# Super-light electromagnetic wave with longitudinal and transversal modes

M.M. Kononenko[*]

*Institute of Applied Mechanics of Russian Academy of Science
Leninskiy prosp 32A, 119991, Moscow Russia*

**Abstract:** The transformation converting equations invariant under Lorentz into the equations invariant under Galileo is obtained. On this basis: **(1)** the super-light electromagnetic wave with longitudinal and transversal modes is found out; **(2)** it is shown the wave velocity coincides with that of de Broglie's wave; **(3)** the connection between Maxwell's electrodynamics and Shrödinger's equation is established; **(4)** structural elements of space are discovered and "a horizon of visibility" is found.

It is shown Bell's inequalities and the principle of the light speed constancy are based on the SRT artifact and "Einstein's local realism" is determined by the wave referred above.

Objectivity of results for quantum and classical objects is discussed.

*Keywords: Bell's inequalities; Super-light velocity; Space*

## 1. Introduction

At present it is widely accepted, as a result of Bell's theorem and the EPR experiments (e.g., [1]), that "Einstein's local realism" must be rejected. I am going to show that this idea is erroneous. It is a consequence of fallacious understanding of the locality principle that Einstein used.

According to the postulate of the Special Relativity Theory (SRT) velocities of interactions must be limited to the light velocity and this limitation relates to the locality principle. Such an interpretation of the principle, as it is known, is the physical basis of Bell's inequalities. In the present paper it will be shown: **(i)** the interpretation of the principle contradicts to the SRT; **(ii)** the postulate is one of artifacts of the SRT; **(iii)** "Einstein's local realism" is provided with the L. de Broglie wave, [2], that is taken into account by Einstein's theory in an implicit kind.

To execute the program (i) - (iii) we shall represent the relativity in terms of absolute space - time of Newton. As it is obvious, such a representation is possible only if Newton's space - time can be considered as an abstract mathematical concept.

In the theory based on the concept as postulates we have taken: the relativity and locality principles and Maxwell's equations. Not setting properties of space, we suppose that they should be revealed from the known fundamental equations and principles which, on their nature, embody these properties.

As a result of the premises, we have obtained the Lorentz – Galileo Transformation (L-GT), [3-5]. (Here alongside with new results the basic ones of these works will be briefly submitted.) The L-GT converts the equations invariant under Lorentz into the equations satisfying the relativity principle of Galileo-Newton. The transformation is an indisputable fact that, nevertheless, contradicts Einstein's theory that we consider to be formally true. To find out the underlying cause of the contradiction, our theory has accepted only such initial premises that are true in the frameworks of the SRT. This fact allows us to reveal artifacts of the SRT which are in part pointed out in the program (i) - (iii).

---

[*] e-mail: kkononenko@mail.ru

## 2. Lorentz-Galileo Transformation

The Lorentz Transformation (LT) that should be obtained with the help of the absolute space-time concept we write down as follows

$$t = t' / \cos\alpha + x' \tan\alpha , \quad x = t' \tan\alpha + x' / \cos\alpha , \qquad (1)$$

$$y = y', \quad z = z' \qquad (2)$$

Here $|\alpha| < \pi/2$, and the system of units in which the light velocity $c = 1$ is used (for transition to one of customary representations of the LT it is necessary to put $sh\varphi = \tan\alpha$ or $v = \sin\alpha$).

Such a record of the LT is convenient for its geometrical representation as mutually orthogonal oblique coordinates $\{x, t\}$ and $\{x', t'\}$ with equal scales on all axes. The angle between the axes $x$ and $t'$ is equal to $\pi/2$ and so is the angle between the axes $x'$ and $t$; the angles between the same-name axes is equal to $\alpha$ (see Fig. 1). This representation is discussed in detail in our work, [4, 5], and the uniqueness of such a geometrical interpretation is proved ibidem.

Obtain the LT formulas (1) using the geometrical representation. For the radius-vector of an arbitrary point of coordinate systems $\{x, t\}$ and $\{x', t'\}$ with a common point of their origin the following equality is true

$$\boldsymbol{r} = t\,\boldsymbol{i_t} + x\,\boldsymbol{i_x} = t'\,\boldsymbol{i_{t'}} + x'\,\boldsymbol{i_{x'}} \qquad (3)$$

The relative position of the coordinate axes is defined by scalar products of the unit vectors of the axes as follows:

$$(\boldsymbol{i_t}, \boldsymbol{i_{t'}}) = (\boldsymbol{i_x}, \boldsymbol{i_{x'}}) = \cos\alpha, \quad (\boldsymbol{i_x}, \boldsymbol{i_t}) = -\sin\alpha, \quad (\boldsymbol{i_{x'}}, \boldsymbol{i_{t'}}) = \sin\alpha, \quad (\boldsymbol{i_x}, \boldsymbol{i_{t'}}) = (\boldsymbol{i_{x'}}, \boldsymbol{i_t}) = 0 \qquad (4)$$

After scalar multiplication of Eq. (3) by $\boldsymbol{i_{t'}}$ and by $\boldsymbol{i_{x'}}$, taking into account Eqs. (4), ones obtain two equalities that coincide with LT (1).

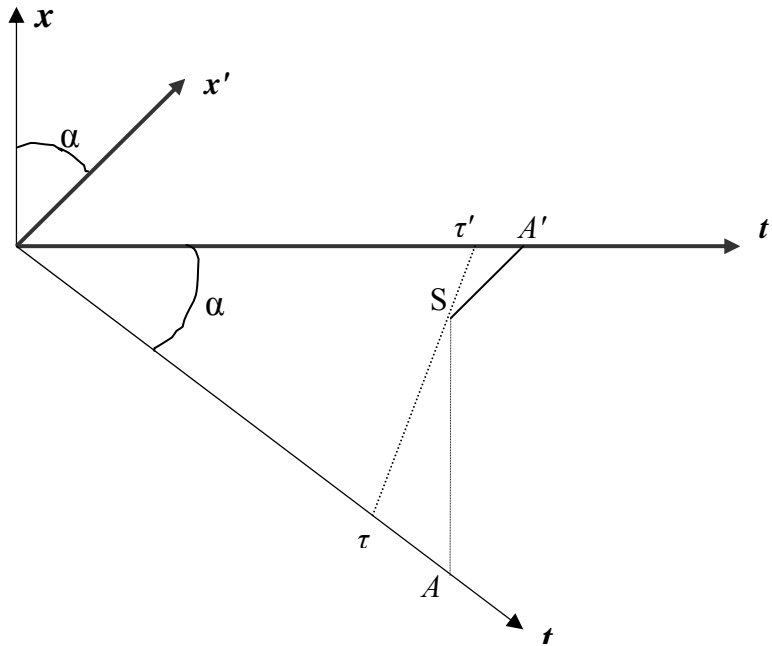

Fig. 1. Geometrical representation of the Lorentz transformation (1). $S$ is the event at the moment of absolute time $\tau = \tau'$; $SA = x$ and $SA' = |x'|$ is the coordinates of the event in systems of readout $K$ и $K'$, respectively.

As it usually is, the non-primed system of coordinates is bound up with an observer who is located at the origin of coordinates of an inertial reference system $K$. The primed system is bound up with a material particle located at the origin of coordinates of another inertial reference system $K'$ moving with respect to $K$ uniformly and in straight lines along the axis $x$. It is more correctly to speak about an inertial configuration $KK'$ connecting two material particles with each other, but it is possible to use habitual terminology taking into account, however, that reference frames have no physical properties.

The velocity of $K'$ with respect to $K$ should be expressed through the parameter $\alpha$. By the SRT definition, the system of readout $K'$ determined in the LT by the condition $x' = 0$ moves along the axis $x$ with the velocity $v = x / t = \sin \alpha$. The definition seems to be evident. However, it is necessary to perceive that it extends the velocity understanding based on kinematics of Galileo to the case when this kinematics is rejected.

A correct velocity definition can be obtained only by means of abstract concepts of temporal and spatial coordinates, for in that case the theoretical tool cannot be a source of artifacts distorting a picture of reality. Such a definition is given by the Galileo Transformation (GT):

$$\tau = \tau', \quad x = u\tau' + x' \tag{5}$$

Here $u$ is the normalized velocity (in the system of units in which $c = 1$); $\tau$ is the absolute time; the non-primed and primed coordinates are connected, correspondingly, with $K$ and $K'$.

Let's notice such a velocity definition demands to model the ideal concepts of time and length by means of devices. However this requirement is not additional in comparison with SRT which, as it will be shown, also requires these ideal measures.

The fact of a physical interaction of two material bodies means that a change of the physical state of the first body entails the corresponding change of the second. Concretizing the locality principle, as it is accepted in [6], let's assume that there is a signal in Nature which transfers information about parameters of such an event from the first body to the second. The hypothetical signal, on its functional role, can be named as the "correlation signal" or briefly "$B$-signal".

This conception can be generalized to any events occurring in physical space. Let $(\tau, x)$ and $(\tau', x')$ be coordinates of an event in the $K$ and $K'$, respectively. The event has taken place on the $x$-axis segment connecting the coordinate origins of these systems (see Fig. 1), and $\tau = \tau'$ is the absolute time of the event. Such a choice of the event place is not accidental; in work [5] it has been shown that at other locations the causality principle is broken (the registration time of the event precedes the event time).

Suppose that the $B$-signal velocity is equal to $b = const$ for the given situation. In that case we obtain the time instants, $t$ and $t'$, when the signals are registered by observers in the points of origins of $K$ and $K'$, respectively, in the form

$$t = \tau + x / b, \quad t' = \tau' - x' / b \tag{6}$$

Here it is taken into account that for the chosen event $x > 0$ and $x' < 0$; relationships (6) are built upon kinematics of Galileo. Below, the time $t$ and $t'$, measured on the same absolute time scale, is named as "the relative time".

Excluding the absolute time, $\tau$ and $\tau'$, from GT (5) by means of Egs. (6), ones obtain the following transformation

$$t = t'(1 + u / b) + x'(2 + u / b) / b, \quad x = t'u + x'(1 + u / b) \tag{7}$$

Its determinant is equal to unity. It is visible the transformation has the structure of LT (1) but depends on two parameters – $u$ and $b$.

The replacement of operators at transition from the non-primed system of readout to the primed system by means of transformations (2), (7) is carried out under formulas

$$\frac{\partial}{\partial t} = \left(1 + \frac{u}{b}\right)\frac{\partial}{\partial t'} - u\,\frac{\partial}{\partial x'}, \quad \frac{\partial}{\partial x} = -\left(\frac{2}{b} + \frac{u}{b^2}\right)\frac{\partial}{\partial t'} + \left(1 + \frac{u}{b}\right)\frac{\partial}{\partial x'} \quad (8)$$

$$\frac{\partial}{\partial y} = \frac{\partial}{\partial y'}, \quad \frac{\partial}{\partial z} = \frac{\partial}{\partial z'} \quad (9)$$

From Maxwell's electrodynamics and the SRT it is known that the space state depends on a motion. Therefore, a velocity definition must be coordinated with Maxwell's equations. In other words, the definition cannot be arbitrary; there should be a dependence $b = b\ (u)$ that can be obtained from the Maxwell equations. These equations may be written thus:

$$\frac{\partial}{\partial t}\boldsymbol{G} + i\nabla\times\boldsymbol{G} = 0, \quad \nabla\cdot\boldsymbol{G} = 0, \quad \boldsymbol{G} = \boldsymbol{E} + i\boldsymbol{H} \quad (10)$$

Here $i$ is the imaginary unit. The time used in these equations, as it is known, is not absolute. Suppose that this time is the relative one introduced above; the sequel confirms the supposition.

To find the dependence $b = b\ (u)$ we write down Eqs. (10) in reference frame $K'$, using operators (8) and (9). Having done this operation, ones obtain two series of formulas for components of the primed complex vector of an electromagnetic field:

$$G'_x = G_x, \quad G'_y = \left(1 + \frac{u}{b}\right)G_y + i\left(\frac{2}{b} + \frac{u}{b^2}\right)G_z, \quad G'_z = \left(1 + \frac{u}{b}\right)G_z - i\,u\,G_y \quad (11)$$

$$G'_x = G_x, \quad G'_y = \left(1 + \frac{u}{b}\right)G_y + i\,u\,G_z, \quad G'_z = \left(1 + \frac{u}{b}\right)G_z - i\left(\frac{2}{b} + \frac{u}{b^2}\right)G_y \quad (12)$$

The same-name formulas from (11) and (12) coincide with each other and with the SRT results if $u$ and $b$ are connected by the following equation:

$$u = \frac{2b}{b^2 - 1} \quad (13)$$

Parametrizing relationship (13) by means of ordinary trigonometric functions, the motion velocity of $K'$ with respect to $K$ and the $B$-signal speed can be written down thus:

$$u = \tan\alpha \quad (14)$$

$$b = 1/\tan(\alpha/2) \quad (15)$$

Then transformation (7) coincides with LT (1). Formulas (11) and (12) result in the known relationships between the primed and non-primed components of an electromagnetic field:

$$G'_x = G_x, \quad G'_y = G_y/\cos\alpha + i\tan\alpha\,G_z, \quad G'_z = G_z/\cos\alpha - i\tan\alpha\,G_y \quad (16)$$

If to start from the LT, the velocity motion would be possible to define as $u = x / t'$. Consequently, the absolute time coincides with the proper time.

The given derivation of the LT proves the compatibility of the absolute time concept with Maxwell's electrodynamics and shows that the concept is the basis of LT.

However, the result obtained contradicts the SRT postulates because $u \to \infty$ when $\alpha \to \pi / 2$, and the $B$-signal speed always is more than unit (at $\alpha \to 0$ it can be arbitrary large).

At the same time, both the theories give the same quantitative description of most experimental results. For example, such are measurements of distances passing by unstable particles for their life-time ($\tau$). Under our theory, the distance is equal to $L_1 = u\tau$, and within the framework of SRT – $L_2 = v\tau / \cos\alpha$. As $u = v / \cos\alpha$, ones obtain $L_1 = L_2$. The similar result is also obtained for the relativistic definition of a momentum that has a classical form, $p = mu$, coincided with the SRT definition, $p = mv / \cos\alpha$, where $m$ is the rest mass.

Let's obtain the L-GT. For this purpose it is enough to replace in LT (1) the relative time, $t'$, by the absolute time, $\tau'$, using the second of Eq. (6). As a result ones obtain the L-GT for an inertial configuration $KK'$:

$$t = \tau' / \cos\alpha + x' \tan(\alpha / 2), \qquad x = \tau' \tan\alpha + x' \qquad (17)$$

It is self-evident that the L-GT also involves Eqs. (2). Here $t$ and $x$ are the parameters of the Lorentz-Poincaré space-time, and $\tau'$, $x'$ are that of Newton's absolute space-time.

From (17) follows the operator expression of the L-GT:

$$\left(\frac{\partial}{\partial t}\right)_x = \left(\frac{\partial}{\partial \tau'}\right)_{x'} - \tan\alpha\left(\frac{\partial}{\partial x'}\right)_{\tau'}, \quad \left(\frac{\partial}{\partial x}\right)_t = -\tan\frac{\alpha}{2}\left(\frac{\partial}{\partial \tau'}\right)_{x'} + \frac{1}{\cos\alpha}\left(\frac{\partial}{\partial x'}\right)_{\tau'} \qquad (18)$$

The second of Eqs. (17) coincides with the $x$-transformation of Galileo. Therefore, spatial coordinates of the LT is expected to be absolute.

In SRT the dependence of spatial coordinates on a motion velocity arises only at using the proper time which, as it was shown, coincides with the absolute time. This is conditioned by the following circumstance. A constant motion is described by three kinematics parameters (time, distance, and velocity) and only two of them are independent from each other. The SRT introduced an absolute measure of velocities by means of the light speed postulate. Hence, another arbitrary measure (named, otherwise, the absolute one and determined by the will of a researcher and by the historical tradition) can be introduced only for one of two other parameters. Within the framework of the SRT the arbitrary measure is introduced either for the time or for the distance. The choice depends on a specific character of interpretation. Thus the SRT also uses absolute coordinates: absolute temporal and relative spatial coordinates or absolute spatial and relative temporal coordinates.

It is easy to check up that the spatial coordinates of the LT are absolute. If it so, the GT should be deduced from the LT. With this purpose we exclude $t$ and $t'$ from LT (1) with the help of Eqs. (6), identifying the absolute spatial coordinates from Eqs. (6) with the corresponding spatial coordinates of the LT. As the result of elementary calculations we obtain the GT.

In the LT the relative time is measured on an absolute time scale. Its physical sense is completely defined by formulas (6). As evident from the formulas, in the LT the temporal coordinates ($t$ and $t'$) of an event are time moments of influence of the $B$-signal caused by the event upon material particles constituting configuration $KK'$. This peculiarity of the temporal coordinates determines the relative time as an attribute of the $B$-signal. It is valid to say that the relative time is an objective parameter characterizing processes of interactions of material particles. The set of such processes or, in other words, the set of configurations $KK'$, having a place in Nature, determines the set of natural times. The absolute time is introduced as a general measure for this set.

### 3. Maxwell-Galileo Equations

Now it is rather simple to obtain electrodynamics in terms of absolute space-time. For this purpose it is enough to replace operators in Eqs. (10) using formulas (9) and (18). Thus we go into system of readout $K'$ and replace the relative time by the absolute one. After simple computations the Maxwell equations are converted into the Maxwell-Galileo Equations (M-GEqs):

$$\frac{\partial \boldsymbol{G}'}{\partial \tau'} + i\,\bar{\nabla}' \times \boldsymbol{G}' = 0\ , \quad \bar{\nabla}' \cdot \boldsymbol{G}' = 0\ , \quad \bar{\nabla}' = \boldsymbol{i}'_x\left(\frac{\partial}{\partial \tau'} + \frac{1}{b}\frac{\partial}{\partial x'}\right) + \boldsymbol{i}'_y\frac{\partial}{\partial y'} + \boldsymbol{i}'_z\frac{\partial}{\partial z'} \qquad (19)$$

The non-primed and primed components of $\boldsymbol{G}$ and $\boldsymbol{G}'$ are connected with each other by relationships (16). The M-GEqs differ from Maxwell's equations by the operator acting along spatial axes.

Mathematically, Eqs. (19) are isomorphic to Eqs. (10), i.e., any solutions of Eqs. (19) are also the solutions of Eqs. (10). It is possible to say that Eqs. (19) represent electromagnetism in $(\tau, x)$-form and Eqs. (10) are so in $(t, x)$-form.

The M-GEqs must be invariant concerning the GT. Indeed, replacing primed operators of Eqs. (19) by formulas (9) and by the following formulas

$$\frac{\partial}{\partial \tau'} = \frac{\partial}{\partial \tau} + \tan\alpha\,\frac{\partial}{\partial x}\ , \quad \frac{\partial}{\partial x'} = \frac{\partial}{\partial x} \qquad (20)$$

which follow from the GT ones obtain the same equations. In this case, as well as at invariance under Lorentz, relationships (16) should be used.

The operator $\bar{\nabla}'$ of Eqs. (19) shows that the LT and the equations invariant with respect to the LT automatically take into consideration the locality principle as an action of the $B$-signal with speed $b$, i.e., Einstein's local realism is determined by the super-light $B$-signal. In the STR the locality principle is realized implicitly by change of the relative time scale.

Let's find the partial solution of Eqs. (19) in the form:

$$\boldsymbol{G}' = \big(F_x(y, z),\ F_y(y, z),\ F_z(y, z)\big)\exp(\,i(k_1 x + \nu\tau)) \qquad (21)$$

Substituting this expression in system of Eqs. (19) ones obtain the following system

$$\nu F_x + \frac{\partial F_z}{\partial y} - \frac{\partial F_y}{\partial z} = 0 \qquad (22)$$

$$\nu F_y - i\left(k_1 + \frac{\nu}{b}\right)F_z = -\frac{\partial F_x}{\partial z} \qquad (23)$$

$$i\left(k_1 + \frac{\nu}{b}\right)F_y + \nu F_z = \frac{\partial F_x}{\partial y} \qquad (24)$$

$$i\left(k_1 + \frac{\nu}{b}\right)F_x + \frac{\partial F_y}{\partial y} + \frac{\partial F_z}{\partial z} = 0 \qquad (25)$$

Considering this system of the equations as algebraic one, we find that the matrix rank for the unknown $F_x, F_y, F_z$ and the augmented matrix rank are equal to 3. To find the solution dependent only on the component $F_x$ we equate to zero the determinant $M = i(k_1 + v/b)[v^2 - (k_1 + v/b)^2]$.

The condition $M = 0$ gives, firstly, ordinary electromagnetic waves. Their velocities at moving off and approaching a wave source are respectively equal to:

$$c_1 = -1/(1 + \tan(\alpha/2)), \quad c_2 = 1/(1 - \tan(\alpha/2)) \qquad (26)$$

These formulas also follow from the wave equation for the light after its representation in terms of absolute space - time.

Secondly, from the same condition we obtain the equality $v/k_1 = -b$ that corresponds to a phase velocity of the $B$-signal. In this case from Eqs. (22)-(25) ones obtain the following equation for $F_x$

$$v^2 F_x + \frac{\partial^2 F_x}{\partial y^2} + \frac{\partial^2 F_x}{\partial z^2} = 0, \qquad (27)$$

Components $F_y$ and $F_z$ are calculated from relationships

$$v F_y + \frac{\partial F_x}{\partial z} = 0, \quad v F_z - \frac{\partial F_x}{\partial y} = 0 \qquad (28)$$

Periodic solutions of Eq. (27) give for complex components of an electromagnetic field the following expressions (the primes are omitted):

$$G_x = F_0 \exp\left[i v\left(\tau - \frac{x}{b} + \frac{k_2}{v} y + \frac{k_3}{v} z\right)\right], \ G_y = -i\frac{k_3}{v} G_x, \ G_z = i\frac{k_2}{v} G_x, \ k_2^2 + k_3^2 = v^2 \qquad (29)$$

Here $F_0$ is an arbitrary constant.

Thus the M-GEqs describe, at least, two stationary waves – the light one and the wave named as $B$-signal.

## 4. On nature of the $B$-signal

The stated theory does not introduce any new axioms. For this reason the $B$-signal is assumed to be known physics under other name and our theory makes more exact its parameters and/or a functional role in Nature. The most probable candidate (owing to its super-light speed) is the wave of L. de Broglie who postulated and presented the stationary wave in the form, [2]:

$$\exp\left[i v_0\left(t - vx/c^2\right)/\sqrt{1 - v^2/c^2}\right] \qquad (30)$$

Here standard designations are used (in our designations $v/c = \sin\alpha$ and $c = 1$).

L. de Broglie defined the phase speed of the stationary wave as $x/t = c^2/v$. This definition is expressed through the relative time and absolute spatial coordinate, i.e., it mixes objective and conventional parameters. For this reason it does not describe a wave transmission correctly. Let's find the phase speed of the wave, proceeding from kinematics of Galileo. With this in mind we replace the relative time by the absolute time, using L-GT (17). As a result exponent (30) becomes equal to

$$\exp\left[i\,\nu_0\left(\tau' - \tan\frac{\alpha}{2}\frac{x'}{c}\right)\right] \qquad (31)$$

From expression (31) immediately follows that the phase speed of L. de Broglie's wave coincides with the $B$-signal speed

$$b = x'/\tau' = c/\tan(\alpha/2) \qquad (32)$$

This fact revels that another name of the $B$-signal is de Broglie's wave. The wave has an electromagnetic nature and its parameters are described by formulas (29). In contrast to the ordinary electromagnetic wave in vacuum the $B$-signal has three modes (two transversal and one longitudinal ones) of the complex vector $\boldsymbol{G} = \boldsymbol{E} + i\boldsymbol{H}$.

Bell's inequalities, basing on the SRT postulate about the light speed, ignore the $B$-signal. For this reason their violation disclosed by the EPR experiments has no relation to the physical locality principle. As was shown, within the framework of the SRT the principle is taken into account with help of the known dependence of relative time scale on the motion velocity. Thus Bell's theorem proceeds from an erroneous interpretation of the SRT.

The EPR experiments can be used for direct confirmation of the $B$-signal existence and for measurement of its speed. Really, after some influence on the first particle of an EPR pair there is an time interval, $\Delta\tau < L/b$, inside which the second particle state does not depend on the new state of the first (here $L$ is the base of measurements).

## 5. The velocity of light and Doppler's Effect

As it is apparent from formulas (26), the dependence of the light speed on the relative velocity $u$ of its source is not the classical addition of velocities. At $\alpha << 1$ we obtain that $c_{1,2} \cong 1 \pm u/2$ because $u = \tan\alpha \cong \sin\alpha \cong \alpha$. Therefore, to test formulas (26) when $u << 1$ it is necessary to increase the measurement accuracy of velocities.

From the physical viewpoint, Maxwell's equations and the M-GEqs. are in conflict with each another despite of their mathematical isomorphism. It needs to be ascertained what system of equations meets conditions of experiments. With this aim in view Doppler's optical effect may be used. Really, if formulas (26) are true, this effect should arisen as a consequence of the same physical principles which determine its acoustic analogue because the light speed is determined by the state of a substance in which the light is propagated and it is similar in this respect to an acoustic signal. The checkout shows this is the case.

Indeed, the classical theory of Doppler's acoustic effect, for a source moving off at velocity $u$, predicts the following relationship between the registered ($\nu$) and intrinsic frequency ($\nu_0$) of a source (of sound or light):

$$\nu = \nu_0/(1 + u/|c_1|), \qquad (33)$$

where $c_1$ is the speed of a signal (light or sound). In the optical case the signal speed depends on the motion velocity and this dependence is expressed through parameter $\alpha$. Using formulas (14) and (26) from relationship (33) ones obtain the known formula of Doppler's optical effect:

$$\nu = \nu_0\left(\frac{1}{\cos\alpha} - \tan\alpha\right) \qquad (34)$$

This formula was found by Einstein. However, his derivation is rather confusing. Indeed, it is based on the following statement: An electromagnetic field phase in the certain point of the four-dimensional space-time does not depend on a choice of a reference frame. This physical statement is correct within the framework of our theory but it is incompatible with the space concept that is actually considered in the SRT. Within the scope of the SRT the notion "the certain point of space-time" conflicts with the following conclusion of Einstein and Pauli, [7], caused by the SRT: Physical characteristics of space have neither positions nor velocities.

At the consecutive interpretation of the SRT, the continuum of points of every reference frame cannot be considered as a mapping of the continual set of points (events) of the physical space-time (owing to dependence of the time scale on a motion velocity). Consequently, the primary image (the physical space-time) in the SRT appears to be uncertain. And this is noted in the given conclusion of Einstein and Pauli.

Thus, it is possible to say that the derivation of Doppler's effect in SRT and its new derivation confirm the key concept of the given theory: the light speed is determined by the space state characterized with parameter $\alpha$. This fact proves correctness of formulas (26). From the last, for one's turn, it can conclude that Maxwell's electrodynamics should be represented as M-GEgs to describe a real physical situation adequately.

The light propagation should have some feature similar to the transformation of a sonic wave into the shock wave when its source velocity exceeds some critical one ($u = 4/3$; then $c_2 > b$). In this case Doppler's effect should be broken. This fact can be used for testing the theory.

## 6. Structure of space

To investigate the space structure ensuing from our postulates let's consider the following situation. Let's assume that from the origin point of all coordinate systems $K_i$ $n$ material particles move in arbitrary directions. Then for the relativistic description of their movement it is necessary to introduce $n$ configurations $K_i K_i'$. Each of these configurations is determined by the pair of systems of axes $\{t_i, x_i, y_i, z_i\}$ and $\{t_i', x_i', y_i', z_i'\}$ so that Eqs. (1) and (2) are carried out at $\alpha = \alpha_i$. The index $i$ has to attach also to the spatial and temporal coordinates. All the reference frames $K_i$ are connected with the same motionless observer, and the reference frame $K_i'$ is attached with the $i$th material particle, ($i = 1, \ldots, n$).

The velocity of the $i$th particle determines the value of $\alpha_i$ for each configuration. In turn, parameter $\alpha_i$ determines speeds of the light and $B$-signal. On the other hand, the two speeds are completely determined by properties of physical space. It is possible, if and only if physical space is a set of $n$ elements and the $i$th element determines velocities $c(\alpha_i)$ and $b(\alpha_i)$. Such an element is the three-dimensional configuration $K_i K_i'$. Physical laws act within the bounds of configurations.

So, we have a new concept of space. According to the concept, space, as a whole, consists of a set of elements. The space element discovered is the configuration $KK'$ connecting two material particles in their binary relations with each another. A descriptive-geometric form of a configuration if a particle moves in straight line is a space-thread. To meet the conservation laws each of two interacting particles should be a source-sink of some substance from which the space-thread is created.

Thus relativistic mechanics does not deal with material points and unbounded continual physical space. Its objects are three-dimensional configurations $KK'$. The end points of configurations are material particles, and its internal points are some physical substance that has all properties of space but it, in contrast to classical space and the SRT, is not a separate physical object.

In a system from $n$ classical particles a quantity of possible elementary configurations, each of which connects only various particles, theoretically can be equal to $N = n(n-1)/2$, and their total sum is equal to $2^N$. For ordinary conditions this number is huge. For this reason it is difficult to assume that at each given moment of time each particle cooperates with all others. In all likelihood, in the given domain of coordinate space there are free particles which are not belonged to any configurations.

As it is accepted to account, Neother's theorem and the invariance of Maxwell's equations concerning Lorentz's group proves uniformity and isotropy of space. Undoubtedly, this opinion can be considered as a correct physical conclusion if its preconditions have solely the objective nature. However, the L-GT proves that it is not so, namely: 1) the absolute space - time is an abstract concept that has only the mathematical nature and it is determined by the subjective will of a researcher; 2) the concept is the basis of any mathematical simulation of physical laws. Therefore, the nature of physical laws has two components – objective and subjective ones. This feature of physical laws entails artifacts that should be comprehended to understand laws correctly. (Generally speaking, the problem of the subjective component of physical laws is difficult. Its discussion goes beyond the scope of the present work.)

Let's notice that at $u > b$ (when $\alpha > \pi/3$) the correlation of states of material particles cannot be provided. Apparently, such configurations cannot exist and they are disintegrated. If it is so, it is possible the two scripts. First, the particle after destruction of the configuration can remain free and, hence, absolutely non-observable. The collision of such a particle with an observable configuration will look as an absolutely unpredictable event. Second, it can remain observably if there is a way connecting the given particle with the researcher via other configurations. Let $K_1 K_1',..., K_n K_n'$ be a sequence of configurations that determines the light way to an observer. In this case ones obtain the following obvious generalization of Doppler's effect

$$v = \prod_1^n v_0 \left( \frac{1}{\cos\alpha_i} - \tan\alpha_i \right) \qquad (35)$$

As it follows from formula (35), the knowledge of the initial and final frequency is not enough for a conclusion about the motion velocity of a light source. There is "a horizon of visibility" (an object becomes non-observable one without instrumentality of other observable objects). The conclusion is of practical significance for cosmology and it can be used to test the given script.

Concerning the space model generated by Maxwell's equations let's notice the following. On a functional role (to provide interaction of material bodies) and by origin (material particles are a source-sink of the substance making space) the space concept is indistinguishably from the concept of a field.

## 7. On a nature of Shrödinger's equation

The thread-like structure of space will have an essential influence on a particle motion if the particle has few threads connecting it with other particles. This supposition is reasonable for micro-particles. If these representations are true, quantum mechanics cannot do without taking into account the thread-like structure of space and finite speeds of interactions, which determines relativistic properties of space.

From this point of view, the greatest interest represents Shrödinger's equation, for the equation is considered to be not relativistic. Besides, the equation being not invariant with respect to the LT and the GT calls into question the relativity principle. Therefore, checking the offered theory, it is necessary to find out whether Shrödinger's equation takes into account relativistic properties of the reality and the relativity principle.

It is known that de Broglie's wave plays an important role in quantum mechanics. We have obtained the wave from electrodynamics. For the reasons we make use of the solution of the M-GEqs that is submitted by formulas (29). The solution determines three components of a complex vector of an electromagnetic field while Shrödinger's equation is written down for a scalar complex wave function. Therefore, the wave function we introduce as $\psi = \nabla \cdot \boldsymbol{G}$ where the divergence of complex vector $\boldsymbol{G}$ is defined in absolute space-time.

Then from solution (29) ones find

$$\nabla^2 \psi = -ik_1\left(k_1^2 + k_2^2 + k_3^2\right)G_x, \quad \frac{\partial \psi}{\partial \tau} = \nu k_1 G_x \tag{36}$$

Excluding $G_x$ from Eqs. (36), we obtain the equation with structure of Shrödinger's equation for a free particle, otherwise, for configuration $KK'$ that is free from interaction with other objects

$$i\frac{\partial \psi}{\partial \tau} + \frac{\nu}{k_1^2 + k_2^2 + k_3^2}\nabla^2 \psi = 0 \tag{37}$$

For calculations of the Laplacian factor ones take a component of the wave vector from solution (29), $k_1 = \nu \tan(\alpha/2)$, $k_2^2 + k_3^2 = \nu^2$, de Broglie's definition of wave-length, $\lambda = h/m \tan \alpha$, and the kinematic relationship between wave-length and its frequency and speed, $\lambda \nu = 1/\tan(\alpha/2)$. In result we obtain the equation:

$$i\frac{\partial \psi}{\partial \tau} + \frac{h \cos \alpha}{2m}\nabla^2 \psi = 0 \tag{38}$$

This equation can be named as the relativistic equation of Shrödinger. It becomes approximately relativistic one and turns into Shrödinger's equation at $\alpha \ll 1$, when $\cos \alpha \approx 1$.

Thus we have confirmed the theory validity by means of the given derivation of the relativistic equation of Shrödinger. Besides, we have obtained the basis for physical interpretation of wave function and some features of quantum phenomena.

So, in terms of quantum mechanics it is possible to tell that solution (29) and Eq. (38) describe so-called "a pure state". For the description of "a mixed state" of a quantum object connected to other objects by means of $n$ threads it is necessary to consider $n$ systems of the M-GEqs or Maxwell's equations taking into consideration sources and currents. Such a physical sense gets the superposition principle of quantum mechanics if to consider that objects of its description are configurations $KK'$.

The thread-like structure of space discovered within the framework of the classical analysis, allows us to explain also such features of the quantum phenomena as a reduction of wave function and wave properties of particles that can be understood as breaking threads and their contact interactions.

The stated representations for the non-observable substance making physical space can have a practical sense only if the substance can be modeling with the help of the ordinary material environment. The derivation of optical effect of Doppler on the basis of its acoustic analogue allows us to hope that such opportunities exist.

## 8. Conclusion

The physical problem that is necessary to analyze "Einstein's local realism" was to bring to light the physical space properties contained in Maxwell's equations and the Lorentz transformation. As is evident from the given work, the problem was rather simple. It transmuted into an extremely involved problem due to the incompatibility tenet of the absolute time with Maxwell's electromagnetism. The formal faultlessness of SRT and its mathematical beauty, by P. Dirac, ensured a reliable protection of the tenet that turned out to be an insuperable hindrance for the adequate understanding of relativistic properties of reality. In turn, the tenet reliably hid artifacts of the SRT.

The SRT effects having only the mathematical nature and being required for completeness of the theory are: 1) dependences of length, time, and mass upon a motion velocity; 2) the light velocity as the superior limit for velocities of motions and transfers of energy; 3) independence of the light velocity on motion of its source.

The most interesting and practically important result of the stated theory is a discovery of a super-light wave with longitudinal and transversal modes of a complex vector $G = E + iH$ of an electromagnetic field. If we are able to generate and detect this wave, we receive, at least, new means of radio communication.

### Acknowledgement

The author is very much obliged to Ju. T. Medvedev for numerous conversations which promoted a deepening of understanding of the stated problems. I am grateful to Andriankin, E.I., Kadantsev, V.N., Malkin, A.I. and Solov'ev, M.A. for discussions of the work results and useful comments. The author feels deep appreciation to academicians L.I. Sedov and I.F. Obraztsov for support of the given work, which they did in their lifetime.